\documentstyle[preprint,aps]{revtex}
\begin{document}
\draft
\title{Reply to the comment on $^{^{\prime \prime }}$Quantum nonlocality for a
three-particle nonmaximally entangled state without inequaltiy$^{^{\prime
\prime }}$}
\author{Shi-Biao Zheng\thanks{%
E-mail: sbzheng@pub5.fz.fj.cn}}
\address{Department of Electronic Science and Applied Physics\\
Fuzhou University\\
Fuzhou 350002, P. R. China}
\date{\today }
\maketitle

\begin{abstract}
This is to reply to Cereceda's comment on $^{^{\prime \prime }}$Quantum
nonlocality for a three-particle nonmaximally entangled state without
inequaltiy$^{^{\prime \prime }}$
\end{abstract}

\vskip 0.5cm

\narrowtext

In a recent paper [1], I show that the following three-particle nonmaximally
entangled state can exhibit quantum nonlocality without inequaltiy

\begin{equation}
\left| \psi _{1,2,3}\right\rangle =\cos \theta \left| +_1\right\rangle
\left| +_2\right\rangle \left| +_3\right\rangle +i\sin \theta \left|
-_1\right\rangle \left| -_2\right\rangle \left| -_3\right\rangle ,
\end{equation}
where $\left| +\right\rangle $ and $\left| -\right\rangle $ are the spin-up
and down states along the z axis, the subscripts 1,2,3 characterize the
three particles. We here assume that $0<\theta <\pi /4.$ Consider the
physical observables E$_i$ and U$_i$ (i=1,2,3) corresponding to the
operators $\stackrel{\wedge }{E}_i$ and $\stackrel{\wedge }{U}_i$ defined by
Eqs. (6) and (7) of Ref. [1]. The Physical quantities E$_i$ and U$_i$ can
take 1 or -1 corresponding to the eigenvalues of $\stackrel{\wedge }{E}_i$
and $\stackrel{\wedge }{U}_i.$ The predictions of quantum mechanics are

\begin{equation}
\text{if }E_1=1\text{ then }U_2U_3=-1;
\end{equation}

\begin{equation}
\text{if }E_2=1\text{ then }U_1U_3=-1;
\end{equation}

\begin{equation}
\text{if }E_3=1\text{ then }U_1U_2=-1.
\end{equation}
Consider a run of measurements, in which predictions (2), (3), and (4) are
verified and $E_1=E_2=E_3=1$ is obtained. According to local hidden theory,
from the result $E_1=1$ and (2) one can conclude that if U$_2$ and U$_3$ had
been measured one should have obtained $U_2U_3=-1$. On the other hand, from
the result $E_2=1$ and (3) one can conclude that if U$_1$ and U$_3$ had been
measured one should have obtained $U_1U_3=-1$. from the result $E_3=1$ and
(4) one can conclude that if U$_1$ and U$_2$ had been measured one should
have obtained $U_1U_2=-1$. This leads to

\begin{equation}
(U_2U_3)(U_1U_2)(U_1U_3)=-1.
\end{equation}
However, according to local hidden theory these elements of reality $U_i$
have values 1 or -1 and thus $U_iU_i=1.$ This will leads to

\begin{equation}
(U_2U_3)(U_1U_2)(U_1U_3)=1.
\end{equation}
contradicting with Eq.(5). We thus have revealed the inconsistency hidden in
the local hidden variable theory. The self-contraction arises from the
assumption that there exists an element of reality corresponding to each U$%
_i $ even when these quantities are not measured and regardless of what is
done to other systems.

Cereceda argued that the above mentioned derivation is not correct [2].
Cereceda claimed that the measurement of $E_1$ does not provide any
information about the values of $U_2$ and $U_3$ separately so that elements
of reality can not be assigned to $U_2$ and $U_3$ based on the EPR's
criterion. In fact, Cereceda's argument is not correct. EPR's definition of
reality is: $^{^{\prime \prime }}$If, without in any way disturbing a
system, we can predict with certainty ({\it i.e.}, with probability equal to
unity) the value of a physical quantity, then there exists an element of
physical reality corresponding to this physical quantity.$^{^{\prime \prime
}}$ [3]. The measurement of $E_1=1$ results in a perfect correlation of $U_2$
and $U_3$ ({\it i.e.}, particles 2 and 3 collapse to the maximally entangled
state of Eq. (11) of Ref. [2]). In this case, one can predict with certainty
the value to $U_2$ ($U_3$) by measuring $U_3$ ($U_2$), without disturbing
particle 2(3). Therefore, under the condition $E_1=1$, there exist elements
of reality corresponding to $U_2$ and $U_3$ just based on the EPR's
criterion. In the case $E_1=E_2=E_3=1$, there exists an element of reality
corresponding to each $U_i$ based on the EPR's criterion. Thus, the proof of
Ref. [1] is correct.

\end{document}